\documentclass[fleqn,usenatbib,useAMS]{mnras}

\usepackage{graphicx}	
\usepackage{amsmath}	
\usepackage{amssymb}	
\usepackage{multicol}        
\usepackage{bm}		
\usepackage{pdflscape}	
\usepackage{lineno} 

\newcommand{\kms}{\,km\,s$^{-1}$} 
\newcommand{\oii}{[\ion{O}{ii}] }

\usepackage[T1]{fontenc}
\usepackage{ae,aecompl}

\usepackage{newtxtext,newtxmath}

\title[ELG Target Selection Validation for DESI]{Validation of Emission-Line Galaxies Target Selection Algorithms for the Dark Energy Spectroscopic Instrument Using the MMT Binospec}

\author[T. Karim et al.]{Tanveer Karim$^{1}$\thanks{Contact e-mail: \href{mailto:tanveer.karim@cfa.harvard.edu}{tanveer.karim@cfa.harvard.edu}}, Jae H. Lee$^{2,3}$, Daniel J. Eisenstein$^{1}$,
\newauthor Etienne Burtin$^{4}$, John Moustakas$^{5}$, Anand Raichoor$^{6}$, Christophe Y\`{e}che$^{4}$
\\
$^{1}$Center for Astrophysics | Harvard \& Smithsonian, 60 Garden St, Cambridge, MA 02138, USA \\
$^{2}$Department of Physics, Harvard University, Cambridge, MA 02138, USA \\
$^{3}$Vertex Pharmaceuticals, 50 Northern Ave, Boston, MA 02210, USA \\
$^{4}$IRFU, CEA, Universit\'{e} Paris-Saclay, F-91191 Gif-sur-Yvette, France \\
$^{5}$Department of Physics and Astronomy, Siena College, 515 Loudon Road, Loudonville, NY 12211, USA \\
$^{6}$Institute of Physics, Laboratory of Astrophysics, \'{E}cole Polytechnique F\'{e}d\'{e}rale de Lausanne (EPFL), Observatoire de Sauverny, 1290 Versoix, Switzerland}

\date{Last updated 2020 July 27; in original form 2020 April 24}

\pubyear{2019}

\begin{document}
\label{firstpage}
\maketitle

\begin{abstract}
The forthcoming Dark Energy Spectroscopic Instrument (DESI) experiment plans to measure the effects of dark energy on the expansion of the Universe and create a $3$D map of the Universe using galaxies up to $z \sim 1.6$ and QSOs up to $z \sim 3.5$. In order to create this map, DESI will obtain spectroscopic redshifts of over $30$ million objects; among them, a majority are \oii emitting star-forming galaxies known as emission-line galaxies (ELGs). These ELG targets will be pre-selected by drawing a selection region on the $g - r$ vs. $r - z$ colour-colour plot, where high redshift ELGs form a separate locus from the lower redshift ELGs and interlopers. In this paper, we study the efficiency of three ELG target selection algorithms -- the final design report (FDR) cut based on the DEEP2 photometry, Number Density Modelling and Random Forest -- to determine how the combination of these three algorithms can be best used to yield a simple selection boundary that will be best suited to meet DESI's science goals. To do this, we selected $17$ small patches in the DESI footprint where we run the three target selection algorithms to pre-select ELGs based on their photometry. We observed the pre-selected ELGs using the MMT Binospec, which is similar in functionality to the DESI instrument, to obtain their spectroscopic redshifts and fluxes of $1054$ ELGs. By analysing the redshift and fluxing distribution of these galaxies, we find that although NDM performed the best, simple changes in the FDR definition would also yield sufficient performance.
\end{abstract}




\section{Introduction}
\label{sec:intro}

The discovery of the accelerated expansion of the Universe (\citealt{Perlmutter98, Riess98}) with the help of type Ia supernovae (hereafter SNIa) was a paradigm shift in the field of cosmology. Along with this, the discovery of CMB acoustic peaks \citep{Boomerang00, Maxima00}, and the persistent lack of evidence of high $\Omega_M$ \citep{Dekel97, Bahcall98, Freedman00, Primack00}, helped establish the $\Lambda$CDM cosmology as the concordance model of the Universe.

Within the $\Lambda$CDM paradigm, dark energy is hypothesized to be the cause of the accelerated expansion, exerting negative pressure to counteract the effects of gravity. A simple way to parametarize dark energy is using the dark energy equation-of-state $w = \rho/p$. A constant value of $w$, e.g. $w = -1$, would indicate a constant dark energy field whereas a time-dependent $w$ would indicate to more exotic type of dark energy e.g. quintessence (\citealt{Peebles88, Caldwell98}).

Understanding the evolution of dark energy over time would have far-reaching implications both in particle physics \citep{Jimenez03} as well as in cosmological parameter estimation such as the $H_{0}$ measurement. Well-known probes such as SNIa and cosmic microwave background (CMB) anisotropy \citep{Planck18} measure the expansion rate of the Universe at two extremities, i.e. at $z \sim 0$ and $z = 1100$ respectively, but they are also influenced by various systematics and yield $H_{0}$ values that are at tension. Thus, intermediate redshift probes, are important in tracing the evolution of expansion rate of the Universe to break this tension. In particular, tracing the evolution in the epoch $0.2 \lesssim z \lesssim 1.65$ is critically important because it is in this epoch when the Universe went from being matter-dominated to being dark energy-dominated \citep{Comparat15}. Therefore, it is necessary to have probes of expansion in these intermediate redshift ranges. 

The baryonic acoustic oscillation (BAO, \citealt{Eisenstein05}) is one such probe that is a characteristic length scale (a cosmological ``standard ruler") imprinted on the matter distribution of the Universe. It refers to the amount of distance sound waves propagated in the baryon-radiation coupled plasma between the Big Bang and the epoch of recombination. The comoving distance of the propagated distance is $150$ Mpc. Thus, by measuring the expansion of this standard ruler as a function of redshift and comparing it to its comoving distance, one can infer the evolution of dark energy. The simple geometry of the BAO feature makes it a robust probe to study dark energy and its accuracy can be estimated reliably using statistical errors due to absence of any significant systematic uncertainties on these scales (\citealt{Vargas14, Ross15}). 

Recent experiments such as the Baryonic Oscillation Spectroscopic Survey (BOSS, \citealt{Eisenstein11}), the extended Baryonic Oscillation Spectroscopic Survey (eBOSS, \citealt{Dawson16}), and the WiggleZ Survey \citep{WiggleZ14} have used the BAO as a probe to constraint properties of dark energy. While the BAO is insensitive to any significant systematics, it requires a large volume to probe the characteristic length scale. As such, next generation experiments such as the Dark Energy Spectroscopic Instrument (DESI, \citealt{Levi13}) will probe really large volumes ($\sim 10^2$ Gpc$^3$ h$^{-3}$) and provide an unprecedented constraint on the dark energy equation-of-state. 

The goal of DESI is to make high precision measurements of the BAO imprints on the large-scale structures (LSS) up to $z \sim 3.7$ as well as measure the growth of structure through redshift-space distortion (RSD, \citealt{DESI16}). DESI will observe roughly $\sim 14, 000$ deg$^{-2}$ covering up to $z \sim 1.6$ in galaxy redshift survey, up to $z \sim 2.1$ in QSO redshift survey and up to $z \sim 3.7$ in QSO Lyman-$\alpha$ survey. It is a multi-object spectrograph that is capable of observing $5000$ targets simultaneously between $3600 - 9800$ \AA\ wavelength range, with a resolution between $2000 - 5500$. 

DESI will be using four classes of tracers for cosmological surveys -- Bright Galaxy Samples (BGS), Luminous Red Galaxies (LRGs), Emission-Line Galaxies (ELGs) and QSOs. With these four classes combined, DESI will obtain redshifts of over $\sim 30$ million objects. In order to obtain sufficiently large samples of spectroscopic targets, each of these tracer classes have their own target selection algorithms and these algorithms use photometric data from the Legacy Survey \citep{LegacySurvey19} as their inputs (more details in Section~\ref{sec:ls}). Pre-selection is done to warrant high efficiency and spectroscopic completeness. Specifically, in different tracer classes we identify spectral features that are expected to produce a reliable redshift determination within the DESI wavelength range. 

Among the galaxy samples, ELGs form the $2/3^{\text{rd}}$ fraction of the total. While BAO experiments such as BOSS have used LRGs, they are good probes up to $z \sim 1$. To go beyond that, we need to use other galaxy tracers that are more numerous and luminous. Because the global cosmic star formation rate (SFR) peaked around $z \sim 2$ \citep{Madau98} and because it was an order of magnitude higher around $z \sim 1$ compared to today \citep{DESI16}, galaxies with high SFR are the natural choice of tracers to probe $1 < z < 2$. ELGs are galaxies that are star-forming and exhibit strong nebular emission lines originating in the ionised Stromgren Sphere region of short-lived, luminous massive stars. The high SFR of ELGs pushes their integrated colour to the bluer side of the spectrum.

In addition, ELGs emit the \oii 3627-3729 \AA\  doublet within the DESI wavelength range. Since the doublet lines are separated by only $2.783$ \AA\ in rest-frame, they serve as a reliable signature to measure spectroscopic redshift of ELGs without having to detect their continuum or other lines. This property is well-suited for large-volume surveys such as DESI because one can obtain a large number of reliable redshifts with a relatively short integration time. 

Ultimately, the goal of the ELG targeting is to produce enough redshifts in the $0.6$ - $1.6$ range to measure the power spectrum above the shot noise limit. An incorrect targetting approach could reduce the total sample size and bias the BAO measurements. In light of this, \citet{DESI16} proposed to implement simple cuts on the $g-r$ vs. $r-z$ colour-colour space to isolate $0.6 < z < 1.6$ ELGs with sufficiently high flux ($8 \times 10^{-17}$ erg s$^{-1}$ cm$^{-2}$) in the \oii doublet, known as the FDR(Final Design Report) cut. 

However, one concern is that the FDR is based on DEEP2 coverage which does not have ELGs beyond $z = 1.3$. Hence, to assess the performance of FDR at really high redshift range, a need for a pilot survey was felt. In addition, the collaboration internally proposed two additional second generation target selection algorithms that are motivated by different principles, e.g. target more low-flux ELGs, to achieve the same goal as FDR. With three different target selection algorithms, it became important to understand and explore how these algorithms are selecting ELGs from the $grz$ parameter space. As a consequence, we conducted a pilot survey of ELGs using the MMT Binospec using these three target selection algorithms.

In this paper, we present the results of this survey to better understand the efficiency of these selection functions, suggest improvements on the selection functions and publish redshift and flux measurements of $1054$ ELGs. In Section~\ref{sec:ls}, we discuss the Legacy Survey photometric catalogue, which will be the source catalogue for DESI target selection. In Section~\ref{sec:ts}, we discuss the three proposed target selection algorithms and their implementations. Section~\ref{sec:bino} discusses how the MMT Binospec Survey was conducted. Section~\ref{sec:method} discusses the data reduction and analysis pipeline. In Section~\ref{sec:result} we discuss the key findings of our work. Finally, in Section~\ref{sec:conc}, we discuss the implications of our results on the DESI experiment, specifically on the Survey Validation phase of the experiment, which is scheduled to start around August 2020. 

\section{The Legacy Survey}
\label{sec:ls}
The DESI Legacy Surveys\footnote{http://legacysurvey.org/} are a combination of three massive multi-year photometric surveys --- the Dark Energy Camera Legacy Survey (DECaLS), the Beijing-Arizona Sky Survey (BASS) and the Mayall $z$-band Legacy Survey (MzLS) --- that jointly covers $\sim 14000$ deg$^{2}$ of the sky visible from the northern hemisphere in $grz$ bands using telescopes at the Kitt Peak National Observatory (KPNO) and the Cerro Tololo Inter-American Observatory (CTIO). DECaLS, BASS and MzLS have single-frame PSF depths of ($g = 23.95$, $r = 23.54$, $z = 22.5$), ($g = 23.65$, $r = 23.08$) and ($z = 22.6$) respectively. In addition, the Legacy Surveys contain WISE and NEOWISE all-sky data covering four bands at $3.4$, $4.6$, $12$ and $22$ $\mu$m respectively.

The Legacy Surveys were designed to provide reliable photometry of potential DESI targets, and as a result, it coincides with the DESI footprint. Existing photometric surveys such as the SDSS and Pan-STARRS1 are too shallow, and the contiguous extragalactic footprint of SDSS is too small \citep{LegacySurvey19}. Sufficiently high-depth photometric surveys such as the DES survey covers mostly the southern sky, far too south for the Mayall telescope at KPNO, where DESI is being commissioned. Thus, the need for deeper photometry and wider coverage in the northern sky were motivations behind the Legacy Surveys. 

Overall, the Legacy Surveys cover two contiguous regions --- the North Galactic Cap (NGC) covering $9900$ deg$^{2}$ and  the South Galactic Cap (SGC) covering $4400$ deg$^{2}$. DECaLS covered the entirety of the SGC and parts of the NGC, yielding a total coverage of $9000$ deg$^{2}$ in $grz$ bands using the DECam at the Blanco telescope in CTIO. BASS used the $90$Prime camera on the Bok telescope at KPNO to cover $5000$ deg$^{2}$ in the NGC in $gr$ bands. MzLS complemented BASS and covered the same $5000$ deg$^{2}$ in $z$ band using the Mosaic-$3$ camera at Mayall in KPNO. The DESI target selection algorithms are built on implementing selection cuts on the colour-colour space, in particular the $g - r$ vs. $r - z$ colour-colour space for the ELGs. Hence, a uniformity of imaging quality between the three optical surveys is ideal. As a result, all of the three surveys have $\approx 300$ deg$^{2}$ of overlap in the NGC in the range $+32^{\circ} < \delta < 34^{\circ}$ to assist in colour transformation and calibration and an additional $\approx 100$ deg$^{2}$ of overlap in the SDSS Stripe $82$ within the SGC for cross-calibration. 

As of 2019, the Legacy Surveys have published their $8^{\text{th}}$ data release\footnote{http://legacysurvey.org/dr8/description/}. However, at the time of observation of the ELG targets presented in this paper, only DR5 and DR6 had been published, consisting of DECaLS and BASS/MzLS data respectively. Hence, we used DR5 and DR6 for the pilot study mentioned in this paper. 

\section{Target Selection Algorithms}
\label{sec:ts}

Overall, the DESI collaboration has proposed three different target selection algorithms to select ELGs for the experiment --- Final Design Report (FDR) cut, number density modelling (NDM) and random forest (RF). To meet the DESI BAO forecasted precision, two key requirements for target selection have been identified: that ELGs have a target density of $2400$ deg$^{-2}$, with at least $1280$ deg$^{-2}$ successfully measured redshifts, i.e. an efficiency rate of $\sim 53.33\%$, and that the ELGs have a minimum total line flux of $10^{-17}$ erg s$^{-1}$ cm$^{-2}$. 

\begin{figure*}
    \centering
    \includegraphics[width = \textwidth]{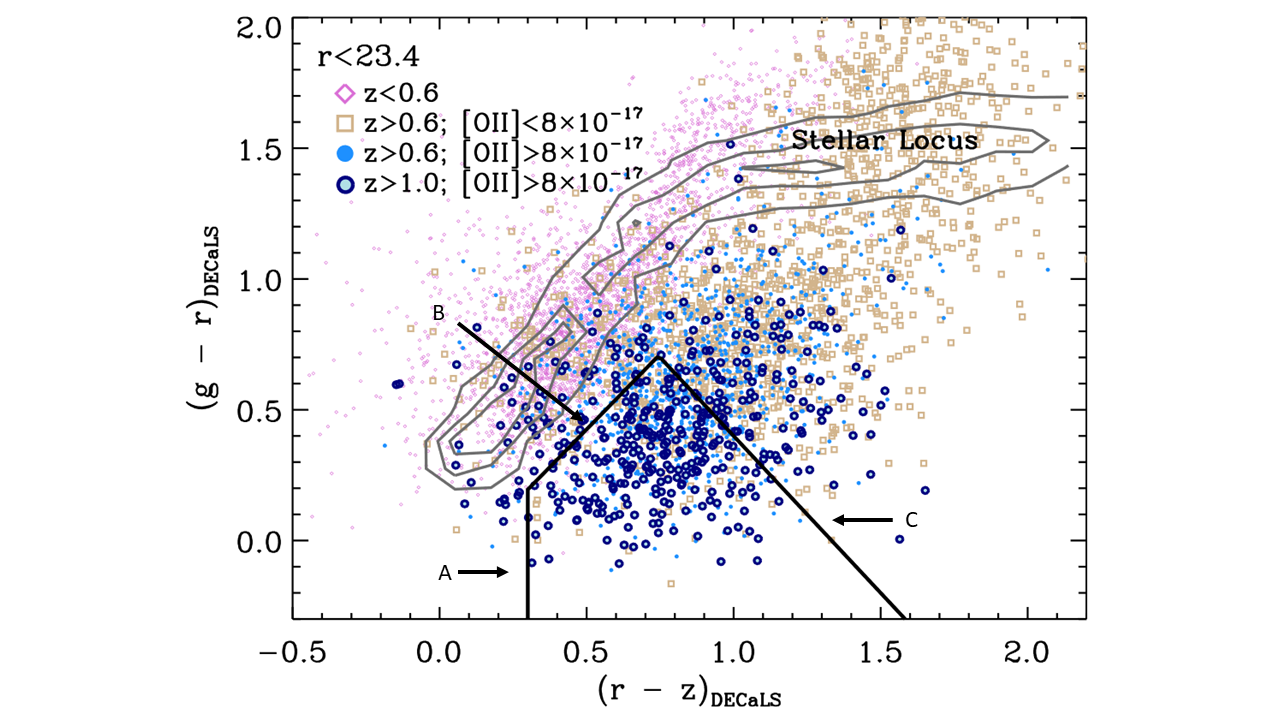}
    \caption{Optical $(g - r)$ vs. $(r - z)$ colour-colour diagram based on spectroscopy from the DEEP2 Galaxy Redshift Survey, illustrating our preliminary selection for ELGs at $z > 0.6$ with significant \oii emission-line flux. Although the galaxy photometry is based on deep CFHTLS imaging \citep{Matthews13}, the colours have been transformed and degraded to the expected depth of the DECaLS imaging. This plot shows that strong \oii-emitting galaxies at $z > 0.6$ (blue points) are in general well-separated from both the population of lower-redshift galaxies (pink diamonds) and from the locus of stars in this colour space (grey contours). The selection box (thick black polygon) selects those galaxies with strong \oii-emission while minimizing contamination from stars and lower-redshift interlopers (Reproduced from \citet{FDR}). We introduce the convention of calling the three cuts as A, B and C as shown in the figure hereafter. The equations of these three cuts are shown in Equation~\ref{eq:fdr}.}
    \label{fig:fdr}
\end{figure*}

In order to meet these requirements, \citet{FDR} proposed to utilize the optical colour-colour parameter space to understand the clustering of the $z \geq 0.6$ ELG population. This strategy was originally implemented by the DEEP2 galaxy redshift survey \citep{Newman13} and has been since used by other programs such as SDSS-III/BOSS \citep{Comparat13} and SDSS-IV/eBOSS \citep{Raichoor17}. 

For DESI, \citet{FDR} specifies that the cuts should be implemented on the $(g - r)$ vs. $(r - z)$ colour-colour space. In this parameter space, the strong ELGs at $z > 0.6$ separate from ELGs at $z < 0.6$ because the spectrum blueward of the Balmer break ($\lambda_{\rm rest} \sim 3700$ \AA) shifts into the $r$-band filter, rapidly reddening the $(r - z)$ colour. Furthermore, at $z \geq 1.2$, the Balmer break moves into the $z$-band filter, causing both $(g-r)$ and $(r-z)$ to be blue at higher redshifts. We observe this phenomena based on the DEEP2 Galaxy Redshift Survey in Figure~\ref{fig:fdr}. Here we notice that the desired ELGs form a cluster near the low-mid part of the plot whereas the low-redshift ELGs form a distinct locus on the top-left part of the plot and stars on the top-right part respectively. 

Thus, the goal of the ELG target selection algorithms is to understand the aforementioned clustering and implement cuts on the $(g - r)$ vs. $(r - z)$ colour-colour space that would separate the desired ELGs from the rest of the population, while meeting the SRD specifications. To do this, we ran a pilot survey using the MMT Binospec and tested the efficiency of the target selection algorithms (details in Section~\ref{sec:bino}). Below, we describe how the three target selection algorithms were designed.

\subsection{Final Design Report (FDR)}

\citet{FDR} proposed a simple polygonal cut on the $(g - r)$ vs. $(r - z)$ colour-colour space shown in black lines in Figure~\ref{fig:fdr} and in functional form in Equation~\ref{eq:fdr}: 

\begin{align}
\label{eq:fdr}
    r &< 23.4 \\ 
    (r - z) &> 0.3 \quad &\text{Cut A} \\
    (r - z) &< 1.6 \\
    (g - r) &< 1.5(r - z) - 0.15 \quad &\text{Cut B} \\
    (g-r) &< 1.6 - 1.2(r - z) \quad &\text{Cut C}
\end{align}

Generally, cut B separates the selection region from low redshift interlopers and cut C separates the selection region from low flux ELGs. The target selection was designed to balance between maximising the ELG density at $z \sim 1$ with s
ignificant \oii flux while minimising low-redshift interlopers and stars \citep{Lee18}. This polygonal cut was specifically designed with the help of the DEEP2 Extended Groth Strip (EGS) region survey \citep{EGS07} data. This was the only region within the entire DEEP2 survey that contained ELGs within the desired DESI redshift range. Note that the selection algorithm is implemented on the DECaLS \citep{LegacySurvey19} colour-colour space. Since the EGS is outside the range of DECaLS, the photometric catalogue of DEEP2 EGS \citep[CFHTLS]{CFHTLS15} initially had to undergo colour transformation in order to match the DECaLS photometry. 

While the FDR cut meets the DESI requirements based on the DEEP2 EGS survey data, the primary source of risk is that the EGS sample consists of only $12,000$ galaxies within $0.6$ deg$^2$ area, out of $53,000$ galaxies within $2.8$ deg$^2$ area of the total DEEP2 survey. Because the EGS area and number of spectra is small, it can be significantly influenced by both shot noise and sample variance and yield a biased redshift distribution. 
Hence, part of the motivation for this paper is to validate the FDR selection with DESI Legacy Surveys photometry and make sure that it is indeed helping DESI meet its goals. 

\subsection{Number Density Modelling (NDM)}
\label{sec:ndm}

\citet{Lee18} proposed the NDM selection algorithm --- a type of Gaussian Mixture Modelling method --- in the context of DESI. The NDM method defines a smooth and continuous selection boundary, in contrast to polygonal FDR cuts, and maximises a user-specified utility function given a target number density and a user-provided error model for the fluxes. NDM achieves this by modelling the number density of the different sub-populations that make up the desired population. 

In order to model the number density of sub-populations, NDM first models the intrinsic number densities of the sub-populations based on DECaLS DR5 and DEEP2 data and then convolves them with a user-specified error model to get the observed density. The intrinsic number densities are parametrised by a Gaussian mixture model that is a function of $grz-$ bands. To estimate the parameters of this mixture model, we used Extreme Deconvolution technique, which is a type of expectation-maximisation algorithm designed to find the maximum likelihood estimators of gaussian mixture models given noisy data. For estimating the parameters, we used Fields 3 and 4 from the DEEP2 survey because these fields have approximately $1$ magnitude additional depth compared to that of DECaLS \citep{Lee18}, making their flux measurements more robust and realiable for traiing purposes. A more detailed discussion on how the intrinsic modelling was done can be found in \citet{Lee18}.

or our user-specified error model, we assumed a normal error model for the observed fluxes, which are used to derive observed densities from the intrinsic ones. The standard deviations of our error models are set by each of the $grz-$band's $5-\sigma$ limiting magnitudes; for our purposes those values are $(g_{\text{lim}}, r_{\text{lim}}, z_{\text{lim}}) = (23.8, 23.4, 22.4)$ respectively.

Afterwards, with the help of the user-specified utility function, NDM maps the observed density on to the $grz$ colour-magnitude space. Finally, NDM uses the utility metric to rank-order the pixels in $grz$ colour-magnitude space and fills them up until the desired target density is achieved. Thus, the filled up pixels in the $grz$ colour-magnitude space define the NDM target selection region. If the pixel size is sufficiently small enough, then the discretized pixels yield a sufficiently continuous and smooth boundary. 

As aforementioned, NDM is trained on the ELGs that intersect between DEEP2 and the DECaLS DR5 survey. Based on the intersecting dataset, $5$ classes of sub-populations were defined: Gold (ELGs with $1.1 < z < 1.6$), Silver (ELGs with $0.6 < z < 1.1$), NoOII (ELGs with $0.6 < z < 0.8$ and without \oii flux measurement), NoZ (ELGs without meaured $z$) and, Non-ELG (Misclassified objects).

For this paper, we implemented three variations of NDM to test its capability of -- (i) selection efficiency and (ii) finding high redshift ($z > 1.1$) ELGs. All of these implementations assume a target density of $3000$ deg$^{-2}$ rather than the DESI requirement of $2400$ deg$^{-2}$. We increased the target density to understand the performance and behaviour of NDM beyond the DESI requirements. 

For NDM1, we set the utility metric weights for Gold, Silver, NoOII, NoZ and Non-ELG to be $1$, $1$, $0.6$, $0.25$ and $0$ respectively. Thus, the fiducial NDM gives equal priority to both high redshift and mid redshift ELGs. These weights are designed for maximum selection efficiency \citep{Lee18}. For NDM2, we double the weight of Gold samples and set it to $2$ and for NDM3, we double to weight of NoZ class to $0.5$. Both NDM2 and NDM3 are designed to maximize the search of $z > 1.1$ ELGs at the expense of overall selection efficiency. Further details of NDM can be found in \citet{Lee18}.

\subsection{Random Forest (RF)}
\label{sec:rf}

In addition to implementing NDM, we implemented a machine learning technique called random forest \citep{Breiman01} that constructs decision trees to understand clustering of data. In contrast to NDM modelling the underlying intrinsic photometry of ELGs, random forest (RF) uses the data explicitly to model the clustering without assuming any intrinsic properties. For the RF technique, we used data available in five bands -- $g$, $r$, $z$, WISE $W1$ and WISE $W2$. The training set consisted of $65000$ low redshift ($z < 0.6$) and $55000$ mid redshift ($0.6 < z < 1.6$) from the Hyper Suprime-Cam (HSC) photometric redshift survey data release 1 \citep{Tanaka18}. All the training data are from two patches in the Fat Stripe 82 region because it overlaps with DECaLS. One patch was used for training and the other for validation in order to reduce the effects of over-fitting. 

Similar to NDM, we implemented three versions of RF to get three different selections. RF1 was tuned to meet the DESI specified target density of $2400$ deg$^{-2}$. All the objects in this selection had a random forest classification probability $p > 0.975$ and $r$-magnitude $< 23.6$. We selected the $r$-magnitude cut to replicate the nominal depth of the DESI survey. RF2 was designed to increase the number density to $3000$ deg$^{-2}$, similar to NDM, to understand the performance and behaviour of RF beyond the DESI requirements. All the objects in this selection had a random forest classification probability of $p > 0.965$ and the same $r$-magnitude cut as RF1. The classification probabilities were selected so to match a given DESI fibre budget and was relaxed a little bit to explore the ELG populations. Finally, RF3 was also tuned to a target density of $3000$ deg$^{-2}$, with the same classification probability and $r$-magnitude limit as RF2, but with an additional constraint on the $g$-magnitude such that $g < 23.55$. Since the $g$-magnitude is a proxy for the \oii flux \citep{Comparat15}, a limit on the $g$-magnitude would allow us to impose flux criterion on the selection function.

\section{MMT Binospec Pilot Survey}
\label{sec:bino}

\begin{figure*}
    \centering
    \includegraphics[width = \textwidth]{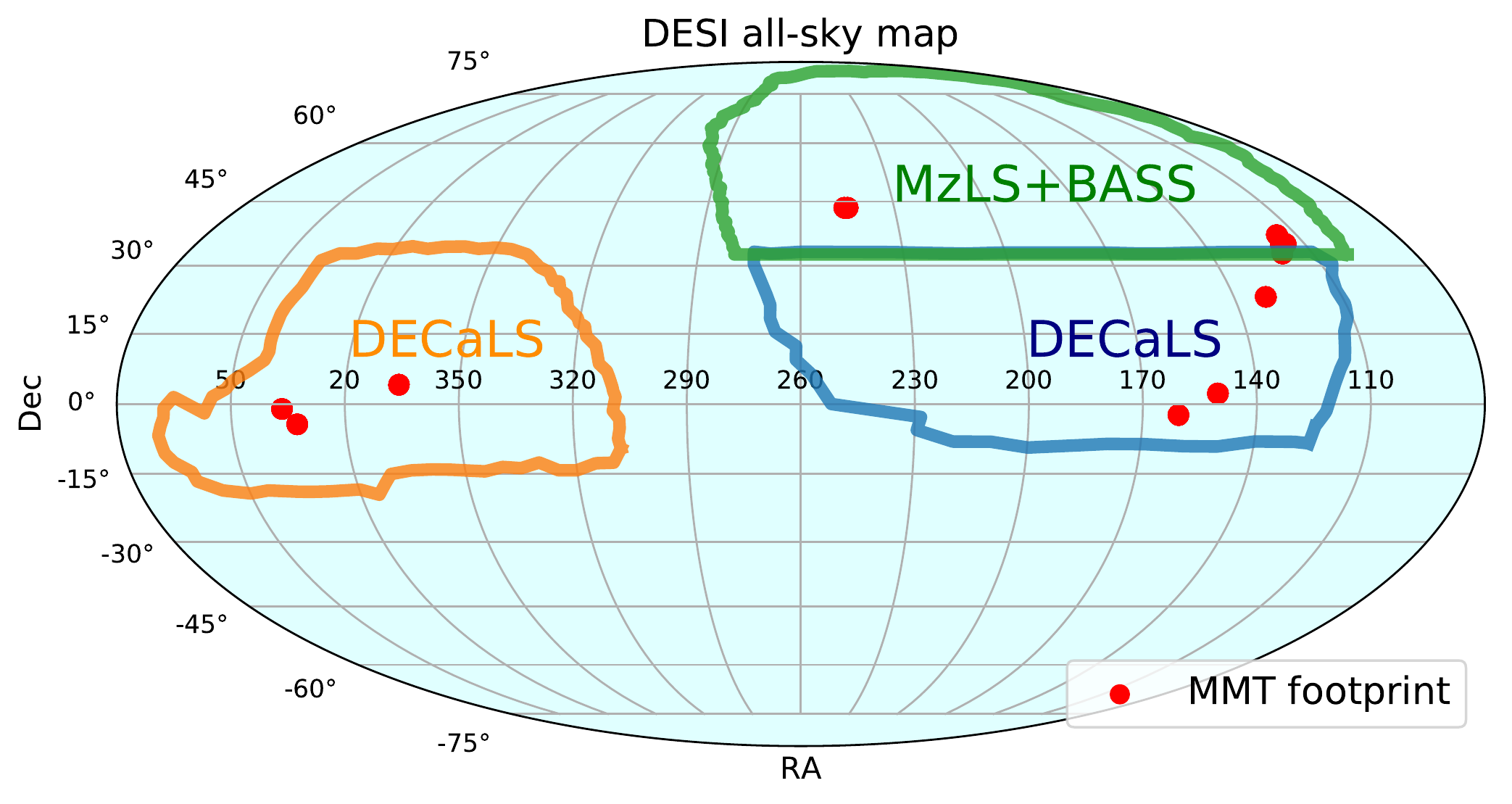}
    \caption{All sky map showing the location of the observed masks and the DESI footprint. The orange and blue dots denote the boundaries of DECaLS and the greend dots denote the boundary of MzLS and BASS. The MMT masks are represented by the red dots.}
    \label{fig:allsky}
\end{figure*}

As discussed in Section~\ref{sec:intro}, we chose to use the Binospec \citep{Fabricant19} at MMT, a slitmask instrument, to conduct our pilot survey because its wavelength coverage and resolutions are similar to that of DESI, ranging between $2000$ and $5500$ over a wavelength range from $360$ nm to $980$ nm. Specifically, we opted to use two $600$ lines/mm grating modes covering the wavelength ranges $4500 - 6960$\AA\ and $7255-9750$ \AA\ with resolutions $2740$  and $4360$ respectively. Hereafter, we call the bluer grating mode ``blue mask" and the redder grating mode ``red mask". All of the ELGs were observed with both blue and red masks. We operated Binospec in maximal slit packing mode, which allowed us to observe $120 - 150$ targets using both masks, each covering $8 \arcmin \times 15 \arcmin$ field-of-view. The slits were of size $1 \arcsec \times 6 \arcsec$. Each field-of-view was observed $3$ times with $10$ minute exposure at every instance, yielding a total exposure time of $30$ minutes per object per grating mode at an expected depth greater than that of DESI. For context, DESI requires imaging depth of at least $24.0$, $23.4$ and $22.5$ AB in $g-$, $r-$ and $z-$ bands respectively with a $5\sigma$ detection of a fiducial galaxy with an exponential light profile of half-light radius $r_{half}$ = $0.45$ \arcsec. Thus, if we had difficulty detecting the \oii line of an ELG, then DESI would certainly not be able to detect it. The greater depth would allow us to probe the limits of DESI and help us understand systematics better. 

We observed a total of $17$ regions over three seasons -- 2017C, 2018A and 2018B. $12$ out of the $17$ regions produced good quality dataset that could be used to determine reliable fluxes of the observed ELGs and we present results from only those $12$ regions. We measured the goodness of dataset by looking at the flux scatter of standard F-type stars from the Sloan Digital Sky Survey \citep[SDSS]{Sloan00} that were included in every mask as flux calibrators. If a mask showed high standard-star flux variance ($\sim 30\%$), we did not include it for our analysis as we were not able to isolate the causes of these high variances, rendering the final flux measurement not useful for our analysis. 

Figure~\ref{fig:allsky} is an all-sky projection map showing the DESI footprint and where our target regions lie; the MMT masks are denoted by red while the other colours represent the DESI footprint. All of these regions are high-valued regions such as COSMOS \citep{Scoville07}, DES \citep{DES16}, HSC \citep{HSC18}, the Southern Galactic Cap and the Northern Galactic Cap. These high-valued regions overlap with different photometric surveys, allowing us to compare the Legacy Survey photometry with that of other surveys as well as avoid any systematics cropping up due to focusing on a particular region on the sky. We assigned equal priority to all the targets, including the standard stars, for ease of analysis.

\section{Methodology}
\label{sec:method}

The methodology process can be broken down into three major parts -- data reduction, determination of spectroscopic redshift, and determination of physical fluxes of ELGs for which successful redshifts were identified. We describe each of these parts in detail below.

\subsection{Data Reduction}
The standard Binospec pipeline\footnote{https://bitbucket.org/chil\_sai/binospec/wiki/Home} outputs a few data products that are useful for redshift and flux measurements of ELGs, i.e. 2D and 1D spectra of the target ELGs. The 2D spectra are co-added, sky subtracted, linearised, corrected for blaze function and pseudo flux-calibrated (not taking sky or slit losses into account). The corresponding 1D spectra are generated from the 2D spectra. However, we wrote our own 1D extraction pipeline because we wanted to have control over the extraction kernel being used to generate the 1D spectra.

To produce the 1D spectra, we cut the 2D spectra into $15$ pixel windows along the position axis (e.g. $y$-axis in Figure~\ref{fig:hyp-test}) with the object being centred in the middle of the window, i.e. the $8^{\text{th}}$ pixel. We determined the size of the spectra cutout window heuristically by determining what the size of the largest ELG was along the position axis. Cutting the 2D spectra into this window allowed us to speed up our computation time without the loss of any critical information. For the extraction kernel, we used the following function:

\begin{equation} \label{eq:extr}
    \exp{\left(-\frac{1}{2}\left( \frac{x - \mu}{\sigma} \right)^4\right)}
\end{equation}

The area of a DESI fibre is given by $A = (\pi/4) (1.5\arcsec)^2 \approx 1.77~{\rm arcsec}^2$, where $1.5 \arcsec$ is the diameter of a fibre. Since for MMT we use slit spectroscopy with a slit width of $1\arcsec$, we need a height of $1.77\arcsec$ to get the same effective area as that of a DESI fibre so that we can compare the flux measurements from our slits to that of DESI's. Thus, the slidth width times the extraction kernel centered at $\mu$, should integrate to $1.77$. Therefore,

\begin{align}
\label{eq:fibre1}
    &\int^{\infty}_{-\infty} \exp{\left(-\frac{1}{2}\left( \frac{x}{\sigma} \right)^4\right)} = 1.77 \\ \label{eq:fibre2}
    &\implies \sigma  = \frac{1.77}{\int^{\infty}_{-\infty} \exp{\left(-\frac{1}{2}x^4\right)}} \\ \label{eq:fibre3}
    &\implies \sigma \approx 0.82
\end{align}

Converted into pixel scale with a plate scale of $0.25\arcsec/{\rm px}$, we get $\sigma = 3.28~{\rm px}$. Note that we initially wanted to use a boxcar function as the extraction kernel but opted to use Function~\ref{eq:extr} because it is reasonably smooth relative to the plate scale of $0.25\arcsec/{\rm px}$. But more importantly, this function allows us to simulate what the DESI fibre would be observing as the effective area of this kernel is equal to that of DESI following the steps of Equations~\ref{eq:fibre1}, \ref{eq:fibre2} and \ref{eq:fibre3}.

\subsection{Redshift Determination}
We used two different methods to determine spectroscopic redshifts of ELGs -- Convolutional Neural Network (CNN) on the 2D spectra, and the Gaussian doublet filter matching on the 1D spectra. 
We implemented both of these techniques due to the different advantages they present. The CNN exploits the entire 2D spectra to identify \oii doublets. This is especially helpful for identifying doublets that may have high velocity and would not appear as nice Gaussian peaks in the 1D spectra. On the other hand, the Gaussian filter matching method is useful as it allows us to measure flux directly as well as get $\chi^2$ statistic from the filter matching process. In addition, using two different methods allow us to cross-validate our redshift measurements. Hence, the both methods complement each other and allowed us to derive robust redshifts for our sample. 

\subsubsection{Convolutional Neural Network (CNN) Method}
\label{sec:cnn}
\begin{figure*}
    \centering
    \includegraphics[width=\textwidth]{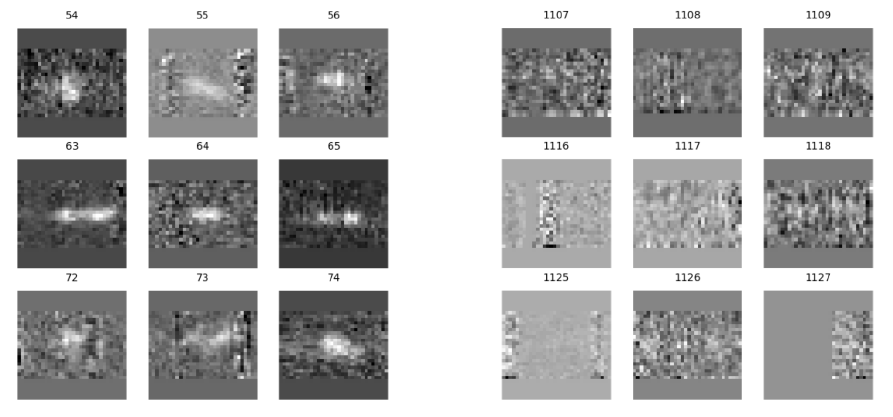}
    \caption{Figure showing examples of ``Blobs" (left panel) and ``Blanks" (right panel). The $x$-axis represents rest-frame wavelength and the $y$-axis represents position axis along the slit. We used these type of cutout stamps of 2D spectra to train the CNN, as described in Section~\ref{sec:cnn}.}
    \label{fig:cnn}
\end{figure*}

A detailed description and implementation of this method can be found in  \citet{Lee18}. For this method, we used a CNN based on the ResNet architechture \citep{He15}. Specifically, we trained our CNN to classify $32 \times 32$ pixel postage stamp cutouts of 2D spectra as either ``Blank" or ``Blob". Examples of such postage stamps are shown in Figure~\ref{fig:cnn}. To train the CNN, we generated $\sim 4 \times 10^{5}$ ``blanks" from various part of the actual 2D spectra where the signal-to-noise was very low and visually inspected to make sure that they are indeed ``blanks". We then added a small amount of random Gaussian noise to these stamps to create an overall $10^6$ number of ``blank" postage stamps. To generate simulated ``blobs", we generated another $10^6$ ``blanks" and then added a mixture of Gaussians with either one or two centroids to mimic real ``blob" examples. With a total of $2 \times 10^6$ training samples, we trained the CNN over $2$ epochs.

After the training was completed, we split up the real 2D ELG spectra into $32 \times 32$ pixel postage stamps. The centre of each postage stamp differs from its neighbour by $5 \times 10^{-5}$ in redshift space. This way, for each ELGs, we constructed a library of possible redshifts where the CNN detected ``blobs". Finally, we visually inspected these libraries to determine spectroscopic redshifts.

\subsubsection{Gaussian doublet filter Method}
\label{sec:gaussfilter}
In the Gaussian doublet filter method, we used the 1D spectra of the ELGs. The redshift of all DESI ELGs will be determined with the help of the \oii doublet, i.e. the $3727$ \AA\ and $3729$ \AA\ emission lines. These lines are very close to each other in rest-frame wavelength, even at $z \sim 1.6$ they appear relatively close to each other; the emission lines appear as two Gaussian peaks (or sometimes as a convolution of the two Gaussian peaks depending on the velocity dispersion and line ratio) in the 1D spectra. To find these peaks, we created a Gaussian doublet model representing the \oii emission lines in the restframe and ran the model as a filter over the entire wavelength space to determine where we see a maximum signal-to-noise. The location of the maximum denotes where the data matches our model the best. By visual inspection, we can verify whether we indeed identified the doublets. The wavelength of the doublets then allows us to easily calculate the spectroscopic redshift of a given ELG.

We defined our model as follows:

\begin{equation}
\label{eq:model}
\begin{split}   
M = \frac{A}{1+r} \frac{1}{\sqrt{2\pi}\sigma^2}\left( r \exp \left[ - \left( \frac{x-\mathcal{\lambda}_{27}(1+z)}{\sqrt{2}\sigma}\right)^2 \right] + \right. \\
    \left. \exp \left[  - \left( \frac{x-\mathcal{\lambda}_{29}(1+z)}{\sqrt{2}\sigma}\right)^2 \right]  \right)
   \end{split}
\end{equation}

\noindent where, $A$, $r$, $\lambda_{27}$, $\lambda_{29}$, $\sigma$ and $z$ represent the amplitude, relative strength ratio between the $3727$ line and the $3729$ line, wavelength of $3727$ line, wavelength of $3729$ line, width of the Gaussians and the redshift at which the model is generated, respectively. Thus, if we fix $r$, we can explore the parameter space as a function of $z$ and $\sigma$ to find where the signal-to-noise (SNR) is maximized. Considering our model $M$ is a Gaussian doublet with values defined in every pixel, we obtain the following:

\begin{equation}
    \chi^2 = \sum_{p} \left( \frac{D_p - M_{p} A}{\sigma_p} \right)^2
\end{equation}
where $p$ stands for pixel in the 1D spectra and $\sigma_{p}$ (not to be confused with $\sigma$ from above) is the flux error in the observed pixel. Taking the first derivative and with simple manipulation, we obtain the following expression for $A$ at extrema:

\begin{equation}
    A = \frac{\sum_{p} \frac{D_p M_p}{\sigma_{p}^{2}}}{\sum_p \frac{M_{p}^{2}}{\sigma_{p}^{2}}}
\end{equation}

The double derivative gives us $\sigma_{A}$ and the expression is:

\begin{align}
    \sigma_{A}^{-2} &= \frac{1}{2} \frac{\partial^2 \chi^2}{\partial A^2} = \sum_p \frac{M_{p}^{2}}{\sigma_{p}^{2}}
\end{align}

Finally, we obtain an expression for SNR which is a function of $M$, and in turn a function of $z$ and $\sigma$ (width of the Gaussian doublets):

\begin{align}
    \text{SNR} = \frac{A}{\sigma_{A}}  &= \frac{\frac{\sum_p \frac{D_p M_p}{\sigma_{p}^{2}}}{\sum_p \frac{M_{p}^{2}}{\sigma_{p}^{2}}}}{\sqrt{\sum_p \frac{M_{p}^{2}}{\sigma_{p}^{2}}}} =  \frac{\sum_p \frac{D_p M_p}{\sigma_{p}^{2}}}{\left( \sum_p \frac{M_{p}^{2}}{\sigma_{p}^{2}} \right)^{3/2}}
\end{align}

For our model $M$, we set $r = 0.7$. The \oii ratio is a function of electron density, ranging from $r = 0.35$ in the high density limit and $r = 1.5$ in the low density limit \citep{Pradhan06}. We opted to choose a value in within the average electron density range without making any assumptions about the astrophysics of the ELGs. To determine at what $z$ and $\sigma$ the SNR is maximized, we simultaneously explored a number of $z$ and $\sigma$ hypotheses. We explored the redshift parameter space from $z = 0.2$ to $z = 1.65$ in steps of $z = 0.0002$. Rather than generating Gaussian doublet models covering the entire wavelength range for every redshift hypothesis, we created smaller windows like the CNN method to speed up the computation process. The centre of each window coincided with the centre of the Gaussian doublet and the widths of these windows were $0.005(1 + z)$. On the other hand, $\sigma$, the width of the Gaussian doublets, is a function of velocity dispersion, $\sigma_v$ and redshift, and its expression is given as:
\begin{equation}
    \sigma = \sqrt{\sigma_{\lambda}^{2} + \sigma_{s}^2}
\end{equation}

\noindent where, $\sigma_{s}$ represents slit aperture and 

\begin{equation}
 \sigma_{\lambda} = \frac{\sigma_v}{c} \lambda_{0} (1 + z)   
\end{equation}
 where $\lambda_{0}$ represents the centre of the Gaussian doublet at rest-frame. For $\sigma_v$, we explored a wide parameter space from $\sigma_v = 0$ \kms~to $\sigma_v = 200$ \kms~in steps of $10$ \kms to obtain the best possible fit. In addition, we set the SNR threshold to be $10$ for successful redshift determination, a little higher than the DESI threshold of $7$ \citep{FDR}, because we did not detect any ELGs  at $7 \leq SNR \leq 10$. 

\begin{figure}
    \centering
    \includegraphics[width = \columnwidth]{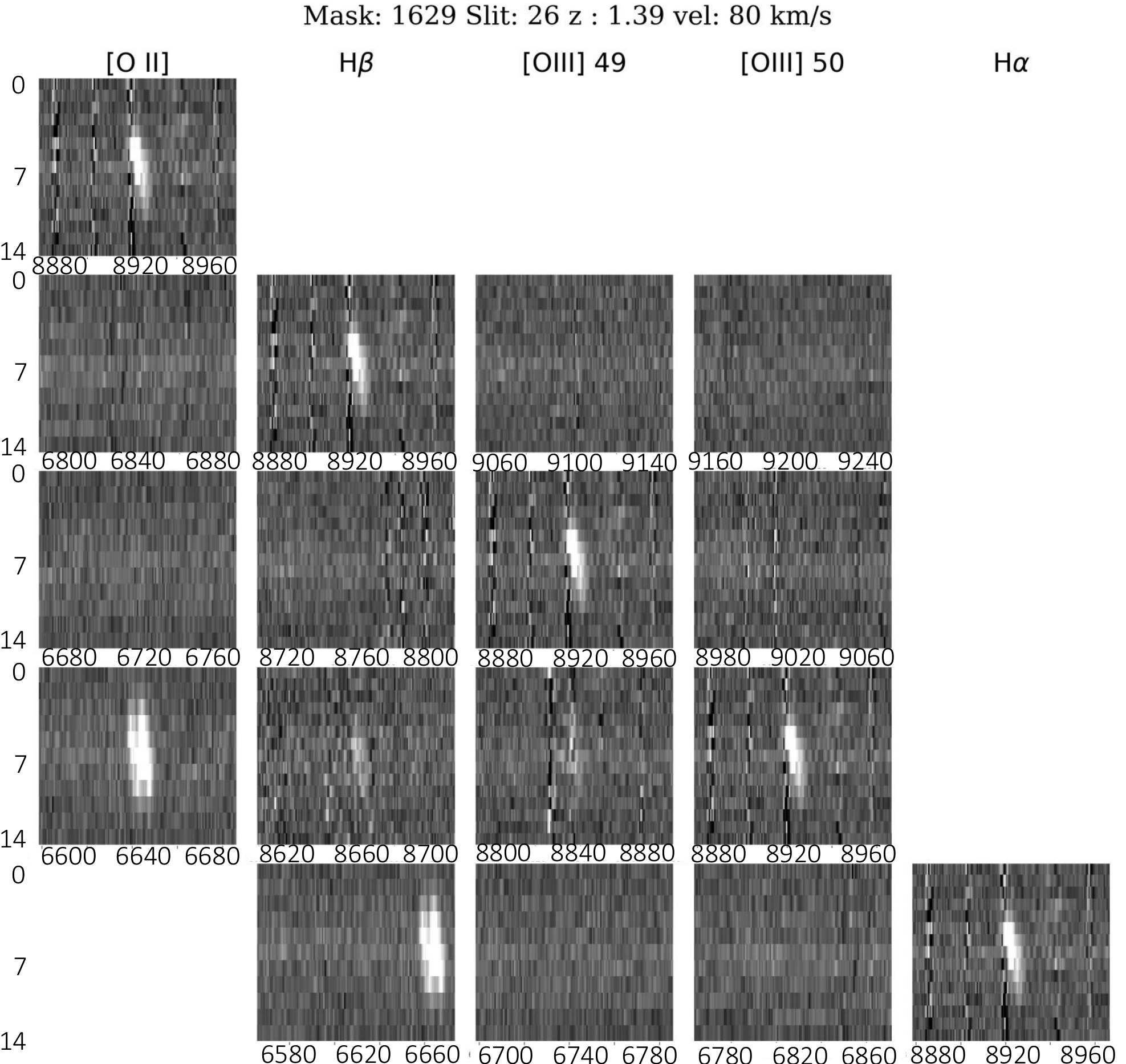}
    \caption{Redshift hypothesis-testing visualization plot of ELG \# $29$ in mask \# $1629$. Our redshift identifier algorithm initially identified a line at $8920$ \AA\ that corresponded to $z = 1.39$. Each row in this plot corresponds to a different hypothesis, with the hypothesis being on the diagonal. For example, the first row corresponds to the hypothesis that the identified line is \oii, the second row corresponds to the hypothesis that the line is $H\beta$ and so on. For the hypothesis [\ion{O}{iii}] $5007$\AA, we see additional lines in the $H\beta$ and \oii panels, confirming that the ELG was given the wrong redshift initially, and the correct redshift should be $z = 0.78$. It should also be noted that hence the velocity of $80$ \kms~is the velocity dispersion of the [\ion{O}{iii}] $5007$ \AA\ line.}
    \label{fig:hyp-test}
\end{figure}

We further visually inspect a given redshift hypothesis to make sure that the identified redshift indeed coincides with a \oii doublet. This is necessary because sometimes blended \oii doublets can appear as a single emission line and for a low-redshift ELG, other emission lines such as the [\ion{O}{iii}] 5007 \AA\ line may be falsely identified as an \oii doublet and we would get the wrong redshift. In order to tackle this problem, we created a visualization tool to check for additional lines for a given redshift hypothesis. 

For example, in Figure~\ref{fig:hyp-test}, we see an example of ELG \# $29$ in mask \# $1629$. Our redshift identifier algorithm initially identified a line at $8920$ \AA\ that corresponded to $z = 1.39$, as shown in the top-left panel. Each row in Figure~\ref{fig:hyp-test} corresponds to a different hypothesis, with the hypothesis being on the diagonal. For example, the first row corresponds to the hypothesis that the identified line at $8920$ \AA is \oii; we then check to see whether given this hypothesis, other corresponding lines ($H\beta$, [\ion{O}{iii}] $4960$\AA, [\ion{O}{iii}] $5007$\AA\ and H$\alpha$) show up in our spectra. We see that for this hypothesis, there are corresponding blank panels for other lines and these blank panels indicate that they are outside the MMT spectrograph's range. Similarly, the second row corresponds to the hypothesis that the same line at $8920$ \AA\ is H$\beta$, but we do not see any corresponding lines which indicates that this line is most likely not H$\beta$. 

For the hypothesis that the line at $8920$ \AA\ is [\ion{O}{iii}] $5007$\AA, we see additional lines in the H$\beta$ and \oii panels, confirming that the ELG was given the wrong redshift initially, and the correct redshift should be $z = 0.78$. The redshift identifier algorithm then goes to the correct redshift and re-fits the Gaussian doublet model to get the correct $\sigma_v$ value.

\begin{table}
\caption{Table defining confidence levels of \oii detection}\label{tab:conf}
\begin{tabular}{| l | l |}
\hline 
Confidence & Description \\
Level & \\
\hline
\hline
$3$ & Distinct \oii doublets  \\
& or high SNR in at least two different lines \\
\hline
$2$ & high SNR in one line \\
& and low SNR in at least another line \\
\hline
$1$ & Detection of only one line (blended \oii line) \\
\hline
\end{tabular}
\end{table}

For visual inspection, we define three levels of confidence on what we can reasonably classify as redshift due to \oii detection. The description of the three levels are given in Table~\ref{tab:conf}. For example, the ELG seen in Figure~\ref{fig:hyp-test} is classified as Confidence $3$ because we see the two lines of the \oii doublet distinctly and can also identify the ELG in two panels with high SNR -- \oii and [\ion{O}{iii}] 5007 \AA. 

Note that, although we measured the redshifts of Confidence $1$ objects with the help of only one blended line, it does not mean that the redshifts obtained for these objects are inaccurate. This is because \oii is the brightest emission line in the blue side of the Balmer break, and if we detect only one line with high enough SNR for a given ELG, then it is most likely the \oii line. We further validate this claim by cross-matching our samples with the HSC photometric redshift catalogue, as explained in Section~\ref{sec:result}.

\subsection{Flux Determination}
\label{sec:fluxing}
Once all the redshifts are determined, we calculate the physical flux of the ELGs with the help of the standard stars previously mentioned in Section~\ref{sec:bino}. The MMT Binospec pipeline outputs the reduced spectra and we need to calculate the calibration vectors in order to convert the reduced spectra into flux. To do this, we use the Python package called SEDPY \citep{sedpy19} that takes SDSS magnitudes (de-reddened) as inputs and generates synthetic stellar spectra as outputs. 

By dividing the synthetic spectra of the standard stars to the observed spectra from Binospec, we can obtain calibration vectors within the Binospec wavelength range. Every mask contains a few standard stars and we take the average of their calibration vectors as the final calibration vector for flux conversion for a given mask. While doing this exercise, we noticed that two of the masks have calibration vectors that differ from each other by a factor of $2$. We could not identify the source of such a large discrepancy but suspect that it is probably due to poor seeing and observing conditions. As a result, we discarded those masks from our analysis. The rest of the masks have calibration vectors that differ up to $20 \%$ in a given mask. Thus, the final fluxes of the ELGs should be treated as having $10 \% - 20 \%$ variation from their true fibre fluxes.

Once the fluxing calibration is completed, we integrate the area under the model \oii for a given redshift and a given $\sigma_v$. The area gives us the aperture \oii flux obtained from an ELG. 

\subsection{Determining Flux Threshold to Simulate DESI-like Observing}
As explained in the key requirements in Section~\ref{sec:ts}, the targetted ELGs will need to yield a minimum line flux of $8 \times 10^{-17}$ erg s$^{-1}$ cm$^{-2}$. However, as explained in Section~\ref{sec:fluxing}, we determined aperture fluxes of ELGs calibrated against point-like sources, i.e. standard stars. Hence, in order to understand whether we have met the minimum threshold flux limit, we need to convert the minimum line flux to aperture flux. Specifically, since our aperture fluxes are calibrated against standard stars, we need to derive the PSF normalized aperture flux value of the line flux.

We derived the PSF normalized aperture flux limit by 
performing rejection sampling. We generate $10^6$ photons within a radius of $10 \arcsec$ with an acceptance probability of $\exp{(-r/b)}$, simulating an exponential profile of a typical ELG. We choose the characteristic length $b = 0.45 \arcsec$, which is the median size of an ELG as per \citet{DESI16}. We find that $\sim 50.13 \%$ of the photons fall within the DESI aperture ($1.5 \arcsec$). We then add the effect of seeing by modelling a 2D Gaussian deviate along $x$ and $y$ position axes with a covariance matrix $\sigma_{xx} = \sigma_{yy} =  1.1 \arcsec/2.35$, $\sigma_{xy} = 0$, and determine what percentage of photons are left within the DESI aperture. Here, $1.1 \arcsec$ is the ideal DESI seeing and $2.35$ is the full width at half maximum. With an average seeing of $1.4 \arcsec$, we find that $\sim 29.32 \%$ photons are left within the DESI aperture. 

Similarly as the ELG, we simulate a point spread function (PSF) with the same 2D Gaussian deviate and find that $\sim 55.38 \%$ of the photons of a simulated star remain within the DESI aperture with a seeing of $1.4 \arcsec$. Thus, the ratio of photon counts of the Gaussian deviated point source and the Gaussian deviated extended source give us a value of $\sim 1.89$, which is the normalizing factor of the desired PSF normalized aperture flux. This implies that the minimum aperture flux threshold for our analysis is $8 \times 10^{-17} \text{erg}~ \text{s}^{-1} \text{cm}^{-2} /1.89 \approx 4.2 \times 10^{-17} \text{erg}~ \text{s}^{-1} \text{cm}^{-2}$. 

\section{Results}
\label{sec:result}

\begin{table*}
\caption{Overall efficiency of different target selection algorithms as a function of confidence class and redshift bins. The numbers represent the percentage of ELGs that meet the criteria mentioned in the columns. For example, the value of the third column in the first row states that $7.2$ \% of the ELGs selected by FDR are in the redshift range $0 < z < 0.6$. The ``After Flux cut" column represents statistics of ELGs whose flux is great than $ 4.2 \times 10^{-17}$ erg s$^{-1}$ cm$^{-2}$.}
\label{tab:sum_stats}
\begin{tabular}{llllllll}
\hline 
Algorithm & Confidence & $0 < z < 0.6$ & $0 < z < 0.6$ & $0.6 < z < 1.1$ & $0.6 < z < 1.1$ & $z > 1.1$ & $z > 1.1$ \\
& Class & & After Flux cut & & After Flux cut & & After Flux cut  \\
\hline
\hline
FDR & $3$ & $ 7.2$ & $1.3$ & $34.0$ & $33.8$ & $11.7$ & $11.5$ \\
& $3 + 2$ & $8.2$ & $1.3$ & $35.9$ & $35.4$ & $12.6$ & $12.3$ \\
& $3 + 2 + 1$ & $8.2$ & $1.3$ & $40.9$ & $40.4$ & $21.6$ & $21.0$ \\
\hline
NDM & $3$ & $1.9$ & $0.2$ & $32.8$ & $32.6$ & $11.4$ & $11.2$ \\
& $3 + 2$ & $2.2$ & $0.2$ & $36.3$ & $35.9$ & $12.1$ & $11.8$ \\
& $3 + 2 + 1$ & $2. 2$ & $0.2$ & $45.9$ & $45.2$ & $22.1$ & $21.2$ \\
\hline
RF & $3$ & $1.0$ & $0.2$ & $28.4$ & $28.4$ & $8.1$ & $7.9$ \\
& $3 + 2$ & $1.2$ & $0.2$ & $32.5$ & $32.3$ & $8.7$ & $8.6$ \\
& $3 + 2 + 1$ & $1.2$ & $0.2$ & $43.4$ & $42.8$ & $16.4$ & $16.0$ \\
\hline
\end{tabular}
\end{table*}

Overall, we obtained reliable spectroscopic redshifts and fluxes of $1054$ ELGs out of a sample size of $1781$ from the $12$ fields. These redshifts are shown in Appendix~\ref{sec:app}, and the catalogue is available as online material. We cross-validate our redshifts with the HSC Public Data Release 2 (PDR2) \citep{HSC19} photometric redshifts. This step was important to make sure that the quality of our redshift measurement of Confidence $1$ ELGs is as robust as our Confidence $3$ and $2$ ELGs because as shown in Table~\ref{tab:conf}, Confidence $1$ objects are limited by observation of only one blended line. We chose HSC PDR2 because its depth ($g \sim 26.6$) is high enough to yield reliable photometry for our sample.

\begin{figure}
    \centering
    \includegraphics[width = \columnwidth]{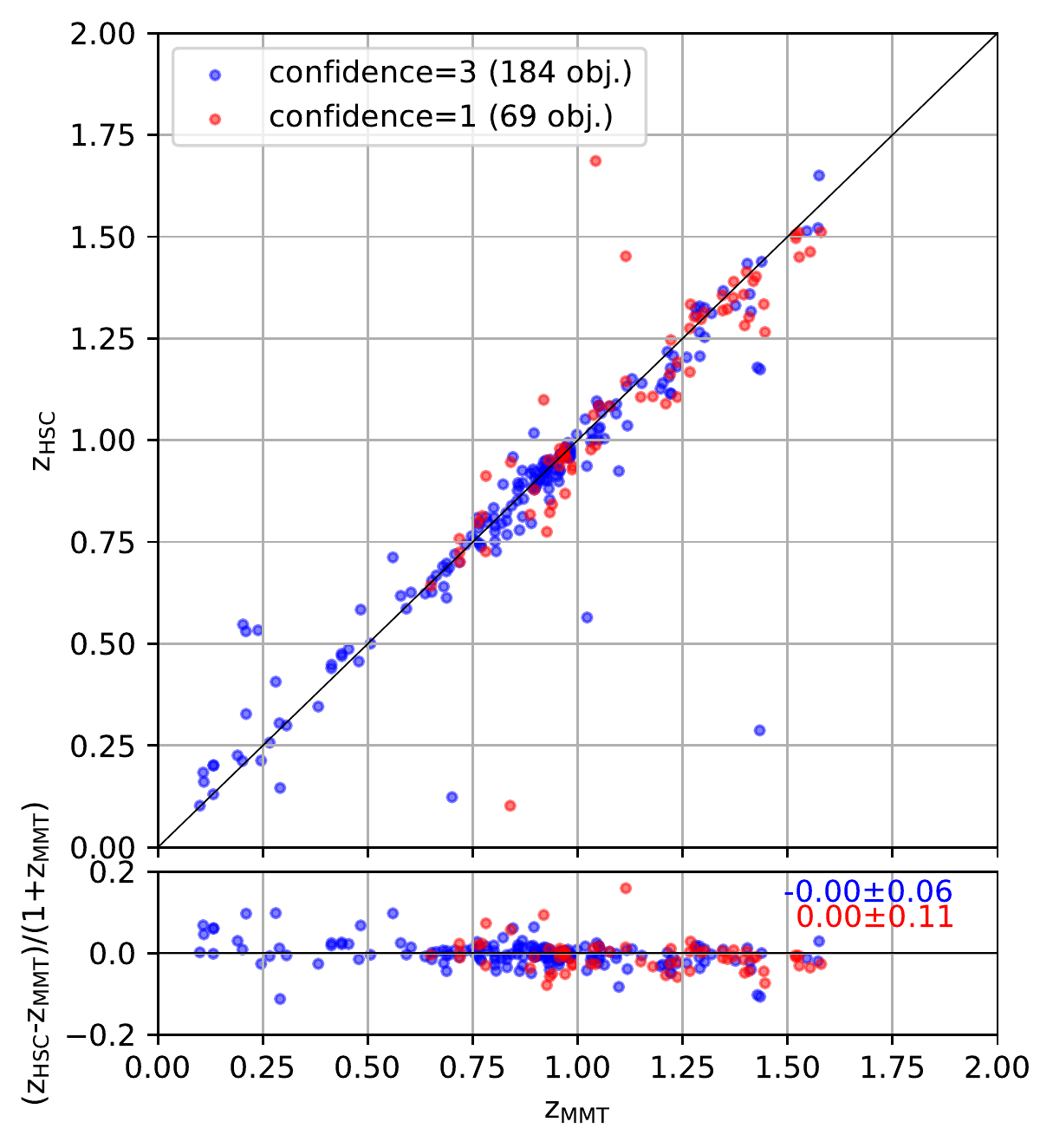}
    \caption{Top: Photometric redshift from HSC PDR2 vs. spectroscopic redshift from this study. Bottom: Redshift residual of HSC PDR2 and this study vs. spectroscopic redshift from this study. Confidence $3$ ELGs are shown in blue and Confidence $1$ ELGs are shown in red. In the bottom panel, the numbers in blue and red correspond to the value of the slopes and the uncertainties for Confidence $3$ and $1$ ELGs respectively.}
    \label{fig:hsc}
\end{figure}

In Figure~\ref{fig:hsc}, we show the photometric redshift from HSC PDR2 vs. spectroscopic redshift from this study, as well the redshift residual vs. spectroscopic redshift from this study for both Confidence $3$ and Confidence $1$ objects. We include the Confidence $3$ objects to show that the HSC PDR2 photometric redshifts are indeed reliable. In fact, the slope of the difference between HSC photometric redshift and the MMT spectroscopic redshift is consistent with $0$ for both confidence classes, indicating that we recovered accurate redshifts for Confidence $1$ objects. This further indicates that the single blended lines we observed are indeed \oii. There are a few outlier redshift measurements but their number is low compared to the overall sample, indicating that on average our spectroscopic measurement can be validated by HSC photometric redshift measurements.

As a result, we are able to consider Confidence $1$ objects in calculating the efficiency of different target selection algorithms. Table~\ref{tab:sum_stats} shows the efficiency of the target selection algorithms as a function of confidence class and redshifts. The numbers represent percentage of ELGs that meet the criteria mentioned in the columns. For example, the value of the second column in the first row states that $7.1$ \% of the ELGs selected by FDR are in the redshift range $0 < z < 0.6$. 

Based on this table, we see that FDR selected the majority of the low-redshift interlopers in contrast to NDM and RF. However, most of the interlopers disappear after the flux cut, not because of low flux, but due to the fact that they are extremely low redshift and the \oii doublet was outside our spectroscopic range to get a flux measurement. For all target selection classes, the mid-redshift range represents the bulk of the ELGs. When we combine all the confidence classes, we see that NDM performed marginally the best among all three with FDR being a close second. %

Also note that the targets observed by the MMT are not all the targets that were chosen by FDR, NDM and RF, i.e. they the targets presented in this paper are a sample from the union of the three algorithm catalogues. Rather than assigning any priority to ELGs that may have been selected as targets by more than one algorithm, we gave all of the targets in the union catalogue equal weight and chose targets randomly. Thus, while we did not observe every target that could have been observed, the targets in this paper nevertheless form a representative sample since they were chosen without any priority consideration.

\subsection{Number of ELGs as a function of redshift}

\begin{figure*}
    \centering
    \includegraphics[width = \textwidth]{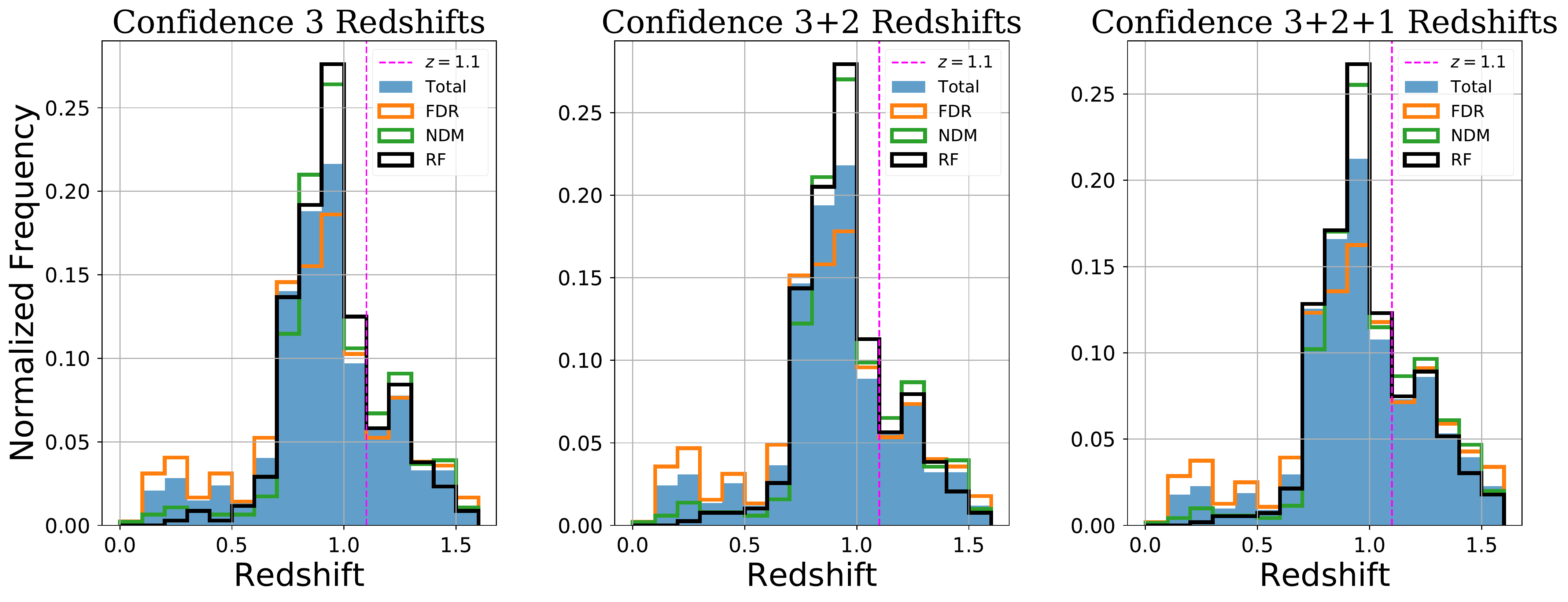}
    \caption{Normalized redshift histogram of ELGs as a function of different selection functions. The $y$-axis represents the normalized frequency in each bin. For example, a bin with a normalized frequency of $0.5$ will contain $50\%$ of the ELGs. Each bin has a width of $\Delta z = 0.1$. The filled blue histogram represents the entire $1054$ population for which we obtained reliable redshifts. Orange, green and black represent FDR, NDM and RF selected ELGs respectively. The magenta line represents $z = 1.1$.}
    \label{fig:redz_hist}
\end{figure*}

We see in Figure~\ref{fig:redz_hist} that the normalized redshift histogram of all the ELGs peaks around $z = 0.9 - 1.0$ with a majority of ELGs lying in the range $0.6 < z < 1.0$. The low redshift range ($0 < z < 0.6$) selection is dominated by FDR, with a higher density than the overall histogram. This would initially suggest that the B cut in Figure~\ref{fig:fdr} should be pushed more towards the lower-right side. However, as we will show in Section~\ref{sec:col-col}, this issue can be attributed to the difference in photometry between DR5/DR6 and DR8 of the Legacy Surveys.

In the medium redshift range ($0.6 < z < 1.1$) both the NDM and RF dominate, with both the NDM and RF proposing over $25\%$ of successful ELGs in the $0.9 - 1.0$ redshift bin. \citet{DESI16} forecasted that the redshift bin that will be sampled the highest is $0.7 - 0.8$ (See Figure 3.11 in \citet{DESI16}). Hence, it is promising to see that any final target selection algorithm inspired by either the NDM or RF will yield more sampling of ELGs at higher redshift bins, enabling DESI to constrain cosmology better at those redshift bins. Finally, in the high redshift ($z > 1.1$) range, we see that NDM performed the best per redshift bin, except at the highest bin $1.5 - 1.6$ where FDR performed the best. Overall, since our pilot survey has more sky area coverage along different patches of the DESI survey compared to the DEEP2 EGS photometry that was based on one region of the sky, our results are more robust and validate the target selection algorithms more uniformly.

As a consequence, the redshift histograms suggest that the second-generation target selection algorithms (NDM and RF) exceeded the nominal efficiency expectations as outlined in Section~\ref{sec:ts}. 

\subsection{Number of ELGs as a function of flux}

\begin{figure*}
    \centering
    \includegraphics[width = \textwidth]{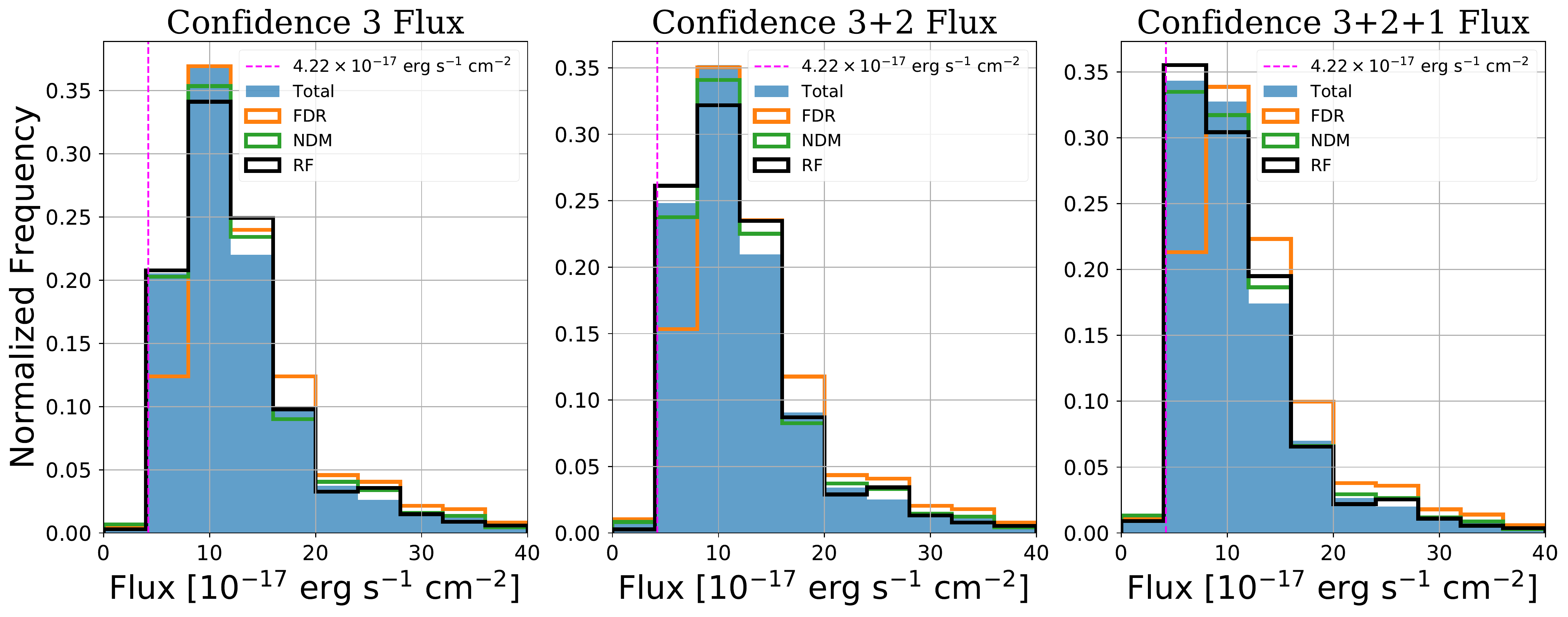}
    \caption{Normalized flux histogram of ELGs as a function of different selection functions. The $y$-axis represents the normalized frequency in each bin. For example, a bin with a normalized frequency of $0.5$ will contain $50\%$ of the ELGs. Each bin has a width of $\Delta f = 4 \times 10^{-17}$ erg s$^{-1}$ cm$^{-2}$. The filled blue histogram represents the entire $1054$ population for which we obtained reliable fluxes. Orange, green and black represent FDR, NDM and RF selected ELGs respectively. The magenta line represents flux $ = 4.2 \times 10^{-17}$ erg s$^{-1}$ cm$^{-2}$.}
    \label{fig:flux_hist}
\end{figure*}

The majority of the ELGs have fluxes in the bin $8 - 12 \times 10^{-17}$ erg s$^{-1}$ cm$^{-2}$ with the bin $4 - 8\times 10^{-17}$ erg s$^{-1}$ cm$^{-2}$ being a close second, as seen in Figure~\ref{fig:flux_hist}. Among the $1054$ ELGs, only $15$ ELGs are below the flux limit threshold of $4.2 \times 10^{-17}$ erg s$^{-1}$ cm$^{-2}$, denoted by the dashed vertical magenta line in Figure~\ref{fig:flux_hist}. This indicates that all the target selection algorithms did well at not selecting the low-flux contaminants, which corresponds to cut C in Figure~\ref{fig:fdr} and Equation~\ref{eq:fdr}.

At the really high flux bins, we see that FDR outperforms both NDM and RF. Interestingly, when we do not consider confidence classes, i.e. the last panel in Figure~\ref{fig:flux_hist}, we see that both NDM and RF sampled more evenly from the two highest flux bins, whereas FDR sampled almost $15\%$ more from the $8 - 12 \times 10^{-17}$ erg s$^{-1}$ cm$^{-2}$ in comparison to the $4 - 8\times 10^{-17}$ erg s$^{-1}$ cm$^{-2}$ bin. This result suggests that that FDR is far more conservative than both NDM and RF along cut C.

Note that, by construction Confidence $1$ objects are those which either do not have resolvable \oii lines in the $2$D spectra or those for whom we could not find additional lines that would verify that the detected lines are indeed \oii. Hence, any ELGs with $z > 1.1$ have a higher probability of being classified as Confidence $1$ because all the additional lines are beyond the spectograph's range. Thus, the shift of NDM and RF from redshift bin of $8 - 12$ to $4 - 8$ when we consider Confidence $1$ objects is promising because this result suggests to us that both of these algorithms are able to select high redshift ELGs that are around the flux threshold.

\subsection{Trend on the colour-colour plot as a function of selection algorithms}
\label{sec:col-col}

\begin{figure*}
    \centering
    \includegraphics[width = \textwidth]{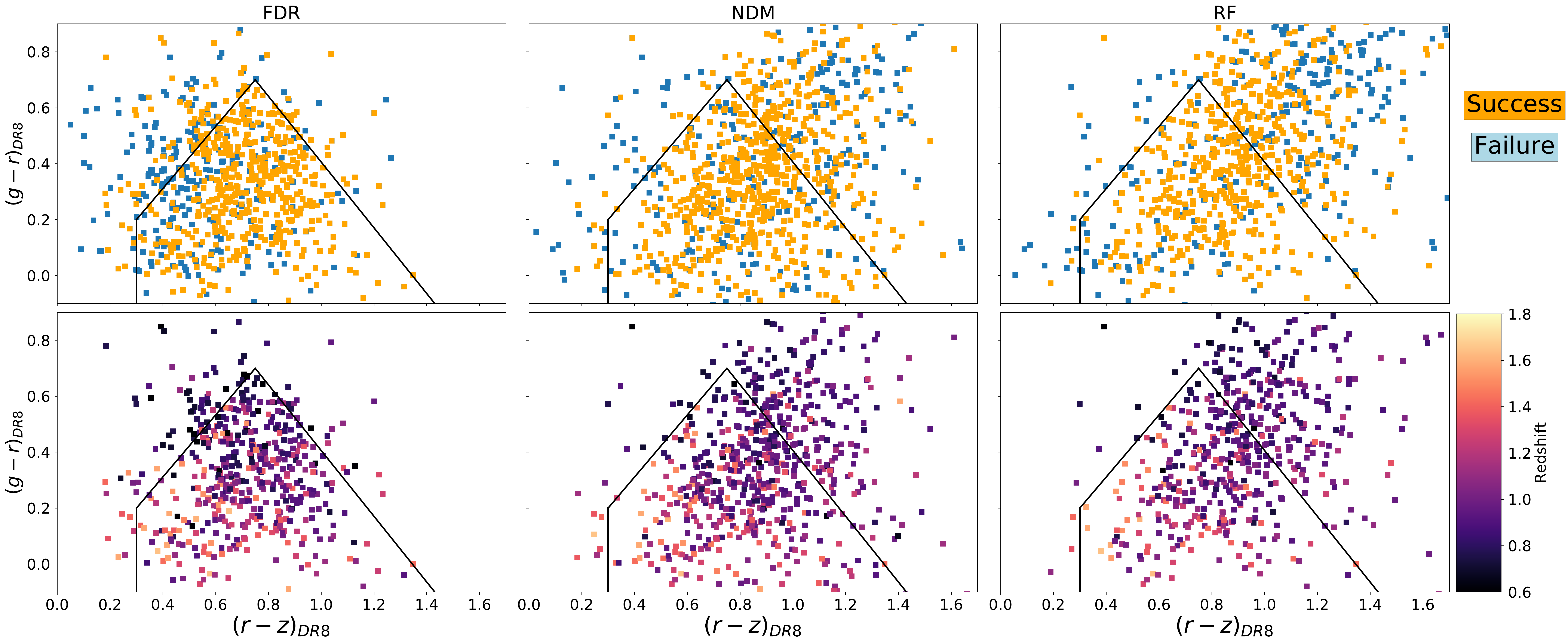}
    \caption{$(g-r)$ vs. $(r-z)$ colour-colour plot of Binospec targetted ELGs. Top row: colour-colour plot as a function of target selection algorithms and success/failure. The colours here are obtained from DR8 of the Legacy Surveys. The orange dots, blue dots, and black lines represent ELGs with reliable redshift and flux determination, ELGs without reliable redshift and/or flux determination and the FDR cut respectively. Bottom row: colour-colour plot as a function of target selection algorithms and redshift distribution represented by the colour bar as the third dimension. The colour bar has a floor value of $0.6$ and a ceiling value of $1.8$, meaning any ELGs with redshift smaller than $0.6$ will be given black colour and any above $1.8$ will be given yellow colour.}
    \label{fig:col-col}
\end{figure*}

Ultimately, the DESI ELG target selection is based on the $(g - r)$ vs. $(r - z)$ colour-colour plot and so, we look at the distribution of success and failure rate as well as redshift distribution of the target selection algorithms in Figure~\ref{fig:col-col}. Specifically, we look for trends that are beyond what the algorithms predicted. The top row of Figure~\ref{fig:col-col} shows the colour-colour plots as functions of success and failure. Each dot represents an ELG; orange corresponds to ELGs for which we obtained reliable redshift and flux measurements, and blue corresponds to ELGs for which we did not obtain redshift or flux measurements. The bottom row shows the colour-colour plots as functions of redshift distribution. In all the panels, we show the FDR cut with black lines. All the colours shown here are from DR8 of the Legacy Surveys. We compare the overall success of each selection algorithm by looking at their individual success (orange) and failure (blue) spread. 

We notice in the top-left panel that the proposed FDR ELGs go beyond the FDR selection. This indicates that there is a difference in data values between DR5/6 and DR8. This result is not surprising because, as explained in Section~\ref{sec:ls}, DECaLS and BASS/MzLS were reduced differently between DR5 and DR6 and as a result, we had to apply our own linear colour-term corrections to convert DR5 to DR6. Any significant non-linearity would introduce error in colour measurement. In contrast, DR8 data are uniformly reduced and colour-corrected, yielding reliable colour measurements. Thus, we attribute the shifting of the FDR ELGs to the systematic differences between DR5 and DR6. 

Interestingly, this bleeding out happens more along cut B, which separates the low-redshift interlopers from our desired ELG population. The significant number of low-redshift interlopers seen in Figure~\ref{fig:redz_hist} can be attributed to the ELG population lying right above cut B. If we ignore those galaxies and only look at the ELGs inside the FDR cut, we see a higher success rate overall, as well as ELGs that fall within the desired redshift range and the low-redshift interloper problem is alleviated. 

While NDM and RF have a different selection region in comparison to FDR, the scatter in colour between DR5/6 and DR8 as seen in FDR indicates that this systematic problem probably affected NDM and RF to some degree as well. However, as discussed in the previous paragraph, qualitatively it appears that the scatter is higher for bluer objects and since NDM and RF proposed more redder objects on average, we expect the effect of this systematic issue to be less in NDM and RF in comparison to FDR. 

Galaxies above cut B have a higher failure rate and low redshift in both NDM and RF, indicating that the definition of cut B is a good one. This interpretation also applies for cut A. Although objects to the left of cut A have high redshifts, this region in the parameter space is prone to small number statistics and would benefit from observation during the Survey Validation phase. 

In contrast, there are a significant number of ELGs just above cut C that yielded successful redshifts and flux measurements. In addition, these galaxies have a mix of redshift values, indicating that cut C could potentially be pushed further towards top-right to gain more ELGs. This result is of interest in the context of DESI because although NDM nominally performed the best, FDR's performance was a close second. If a simple definition change of cut C can yield similar results as that of NDM, then such a selection function would be preferable due to being computationally less demanding. Whereas NDM tries to define a continuous smooth boundary, a simpler definition such as FDR's can be easier to implement and interpret. Therefore, it would be useful to explore the region above cut C during Survey Validation to better understand where NDM sampled from and study whether a simple shift of cut C could result in comparable results as that of NDM.

\subsection{\oii flux to $g$-band flux ratio as function of selection cuts}

\begin{figure*}
    \centering
    \includegraphics[width = \textwidth]{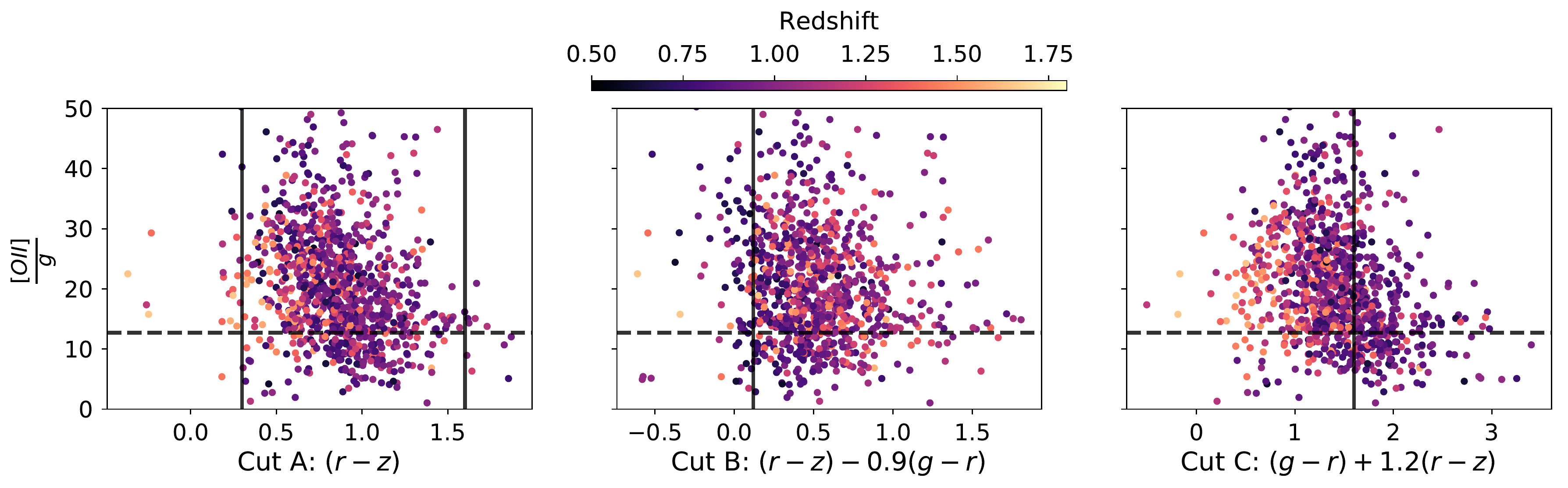}
    \caption{Ratio of \oii flux and $g$-band flux as a function of the three FDR selection cuts and redshifts. The dashed horizontal line is at $12.68$. Objects above this line are DESI observable given a line flux limit of $4.2 \times 10^{-17}$ erg s$^{-1}$ cm$^{-2}$ and a $g$-band magnitude limit of $23.7$. The solid vertical lines define cuts on the colour-colour space. Left: Cut A. The left and right vertical lines represent $0.3$ and $1.6$ respectively. The region bounded by the vertical lines define the selection boundary. Centre: Cut B. The solid vertical line is at $0.12$. Objects with values greater than $0.12$ are inside the selection boundary. Right: Cut C. The solid vertical line is at $1.6$. Objects with values less than $1.6$ are inside the selection boundary.}
    \label{fig:oii-g}
\end{figure*}

In addition to studying the performance of the target selection algorithms, we are interested in understanding how changing observational parameters of DESI could potentially affect selection cuts. We look at the ratio of \oii flux and $g$-band flux versus selection cuts in Figure~\ref{fig:oii-g} to understand this behaviour. Here, we see cut A, B and C in the left, middle and right panel represented by the solid vertical black lines respectively. The dashed horizontal line represents the limiting case where \oii aperture flux is $4.2 \times 10^{-17}$ erg s$^{-1}$ cm$^{-2}$ and $g$-band magnitude limit is $23.7$. Therefore, if DESI imposes a limit $g < 23.7$, then $\sim 80\%$ of the ELGs will be observable. 

The more interesting feature of Figure~\ref{fig:oii-g} are the moderate negative correlations between $\oii/g$ and cut A and cut C, and the absence of any noticeable relationship with cut B. For cut A, we see that pretty much all the objects are at $(r - z) > 0.3$, including almost all of them $(r - z) < 1.6$. While $(r - z) < 1.6$ is an additional cut according to the Final Design Report, we do not give it an explicit name because cut C coincides with this cut at $1.6$. However, from both Figures~\ref{fig:col-col} and \ref{fig:oii-g}, it is apparent that the $(r - z) < 1.6$ can be pushed to $(r - z) < 1.5$ without losing any significant fraction of ELGs. Therefore, DESI can potentially introduce an explicit cut D in Figure~\ref{fig:fdr} as a vertical line at $(r - z) = 1.5$.

On the other hand, this negative correlation relationship is more intriguing for cut C. Since the high redshift ELGs follow the general trend, the high redshift ELGs that are already below the $\oii/g$ limit are unrecoverable. In contrast, if we slide cut C to the right, we will get mostly mid redshift to some high redshift ELGs. Additionally, this analysis suggests that we do not necessarily need to maintain one strict cut C. Rather, we can define cut C as a sliding function of $g$; as $g$ decreases, we can move cut C to the right from the left. We can impose such a definition because $\oii/g$ increases as $g$-band magnitude increases, i.e. $g$-band flux decreases, the dashed horizontal line moves up. 

For example, as seen in Figure~\ref{fig:oii-g}, for $\oii/g \approx 12.68$ it maybe worthwhile to define cut C to be around $2$ rather than $1.6$ because we are losing a large fraction of ELGs. On the other hand, for a depth of $g$ such that $\oii/g \approx 30$, it would be redundant to have cut C defined at $2$; it would make more sense to keep cut C defined at $1.6$ at that $g$-band depth. Hence, we argue that for faint objects, we can reduce the parameter space by sliding cut C to 2. 

\section{Discussion \& Conclusion}
\label{sec:conc}
In this paper, we study the efficiency and performance of three emission-line galaxies target selection algorithms --- final design report (FDR), number density modelling (NDM) and random forest (RF) --- with the help of the MMT Binospec for the forthcoming DESI experiment. We determined redshift and flux of $1054$ ELGs. Our principal findings are as follows:

\begin{enumerate}
    \item There is a significant scatter between colours of DR5/6 and DR8 in the Legacy Surveys, especially for bluer ELGs. This scatter resulted in FDR selecting significantly more low-redshift interlopers. For purely photometry based target selection algorithms, a small deviation in colour measurement can result in non-optimal selection. 
    \item NDM performed the best overall at not selecting low-redshift interlopers. In addition, it performed the best at selecting ELGs in the desired redshift range of DESI, with an overall efficiency of $\sim 66\%$. FDR and RF have overall efficiencies of $\sim 61\%$ and $\sim 59\%$ respectively.
    \item All the target selection algorithms peaked in the redshift bin $0.9 - 1.0$. This redshift bin range is higher than the \citet{DESI16} forecast that suggested the peak to be at $0.7 - 0.8$. Specifically, NDM and RF proposed over $25\%$ ELGs in the $0.9 - 1.0$ bin, signifying that they exceeded the nominal expectations.  
    \item Given an aperture flux limit of $4.2 \times 10^{-17}$ erg s$^{-1}$ cm$^{-2}$, RF sampled $\sim 15\%$ more and NDM sampled $\sim 13\%$ more from the threshold bin of $4 - 8 \times 10^{-17}$ erg s$^{-1}$ cm$^{-2}$, compared to FDR. While FDR and NDM both peaked at flux bin $ 8 - 12 \times 10^{-17}$ erg s$^{-1}$ cm$^{-2}$, RF peaked at $4 - 8 \times 10^{-17}$ erg s$^{-1}$ cm$^{-2}$. Overall, NDM sampled evenly from all flux bins. Therefore, a final target selection cut inspired by either NDM or FDR maybe necessary in order to push DESI to its full limit. 
    \item There are few ELGs around cut A and there is a gap between cut A and where the loci of the three target selection regions are. While it may seem at the first glance that DESI can potentially push the cut A from $0.3$ to $0.4$ without losing too many ELGs, we also find more really high redshift ELGs in this region compared to others. Therefore, it will be paramount to explore the region $0.2 < (r - z) < 0.4$ during the survey validation (SV) phase.
    \end{enumerate}
    
    Overall, the efficiency of FDR ($61 \%$) was close to that of NDM's ($66 \%$), and as shown in Figure~\ref{fig:col-col}, the efficiency of FDR went down due to the scatter in photometric colour for ELGs between DR5/6 and DR8. Ignoring those ELGs and broadening the definition of FDR along cut C would most likely yield similar efficiency as that of NDM. Because the efficiency of FDR and NDM are close, the FDR version of target selection is preferable because FDR uses simple analytic cuts that are easy to interpret, in contrast to NDM that defined a smooth continuous region and can be potentially more difficult to implement. Since we have shown that the risk factors and concerns of using FDR via DEEP2 have been mitigated in this paper, we suggest that an expanded version of the FDR cut be studied during the Survey Validation phase of DESI. Specifically, make the following suggestions:
    
    \begin{itemize}
    \item \textbf{Cut A:} Most of the really high redshift ELGs are close to this cut. Because such high redshift ELGs are the most valuable targets, we recommend exploring $0.2 < (r - z) < 0.4$ more to understand this part of the parameter space better. Currently this region is dominated by small number statistics. 
    \item \textbf{Cut B:} Almost all the objects beyond cut B are either low redshift interlopers or ELGs which did not yield a successful redshift. This result demonstrates that the definition of cut B is most likely a good one.
    \item \textbf{Cut C:} There are many successful ELGs at mid to high redshift range just above C. This region was sampled almost exclusively by NDM and RF. The success of NDM demonstrates the potential of obtaining more successful ELG redshifts by pushing cut C more towards the top-right. While the current definition of cut C is given by $(g - r) + 1.2(r - z) < 1.6$, we recommend to explore up to $2.0$ during the SV phase to better quantify the potential gain by pushing the cut C up. 
    
    We also identified a negative relationship between $\oii/g$ and cut C. This relationship warrants a more in-depth study during the SV phase on the possibility of implementing a sliding definition of cut C, specifically $1.6 < (g - r) + 1.2(r - z) < 2.0$. For bright objects, it may be prudent to consider a wider definition of cut C to observe more ELGs. On the other hand, for faint objects, a wider definition would be redundant. 
    
    \item \textbf{Additional Cut:} Very few ELGs reach up to $(r - z) = 1.6$, with most of them having values less than $(r - z) = 1.5$. This may indicate that an additional vertical cut could be implemented at $(r - z) = 1.5$.
    
\end{itemize}

Our study indicates that photometric target selection is a well-grounded method to sample emission-line galaxies that will yield reliable redshifts. As more and more cosmological surveys will start using ELGs as cosmological tracers, it is becoming of utmost importance to better understand the relationship between properties of ELGs such as their redshift and flux with their photometry in order to construct more robust photometric target selection algorithms. Our study was a test bed for three such algorithms. The results of our work will inform the Survey Validation phase of DESI to define a more robust, efficient and simple target selection region. 

\section{Acknowledgements}
We thank the DESI internal reviewers Chris Miller, Francisco Castander, John Peacock and Ntelis Pierros for their valuable feedback.

TK is supported by the National Science Foundation Graduate Research Fellowship under Grant No. DGE - 1745303. 

DJE is supported by U.S. Department of Energy grant DE-SC0013718 and as a Simons Foundation Investigator.

JM gratefully acknowledges support from the U.S. Department of Energy, Office of Science, Office of High Energy Physics under Award Number DE-SC002008.

Observations reported here were obtained at the MMT Observatory, a joint facility of the Smithsonian Institution and the University of Arizona. 

DESI is supported by the US Department of Energy's Office of Science; the US National Science Foundation, Division of Astronomical Sciences under contract to the NSF's National Optical-Infrared Astronomy Research Laboratory; the Science and Technologies Facilities Council of the United Kingdom; the Gordon and Betty Moore Foundation; the Heising-Simons Foundation; the French Alternative Energies and Atomic Energy Commission (CEA); the National Council of Science and Technology of Mexico; the Ministry of Economy of Spain; and DESI member institutions. The DESI scientists are honored to be permitted to conduct astronomical research on Iolkam Du'ag (Kitt Peak), a mountain with particular significance to the Tohono O'odham Nation.

This research used resources of the National Energy Research Scientific Computing Center (NERSC), a U.S. Department of Energy Office of Science User Facility operated under Contract No. DE-AC02-05CH11231.

The Legacy Surveys consist of three individual and complementary projects: the Dark Energy Camera Legacy Survey (DECaLS; NOAO Proposal ID \# 2014B-0404; PIs: David Schlegel and Arjun Dey), the Beijing-Arizona Sky Survey (BASS; NOAO Proposal ID \# 2015A-0801; PIs: Zhou Xu and Xiaohui Fan), and the Mayall z-band Legacy Survey (MzLS; NOAO Proposal ID \# 2016A-0453; PI: Arjun Dey). DECaLS, BASS and MzLS together include data obtained, respectively, at the Blanco telescope, Cerro Tololo Inter-American Observatory, National Optical Astronomy Observatory (NOAO); the Bok telescope, Steward Observatory, University of Arizona; and the Mayall telescope, Kitt Peak National Observatory, NOAO. The Legacy Surveys project is honored to be permitted to conduct astronomical research on Iolkam Du'ag (Kitt Peak), a mountain with particular significance to the Tohono O'odham Nation.

NOAO is operated by the Association of Universities for Research in Astronomy (AURA) under a cooperative agreement with the National Science Foundation.

This project used data obtained with the Dark Energy Camera (DECam), which was constructed by the Dark Energy Survey (DES) collaboration. Funding for the DES Projects has been provided by the U.S. Department of Energy, the U.S. National Science Foundation, the Ministry of Science and Education of Spain, the Science and Technology Facilities Council of the United Kingdom, the Higher Education Funding Council for England, the National Center for Supercomputing Applications at the University of Illinois at Urbana-Champaign, the Kavli Institute of Cosmological Physics at the University of Chicago, Center for Cosmology and Astro-Particle Physics at the Ohio State University, the Mitchell Institute for Fundamental Physics and Astronomy at Texas A\&M University, Financiadora de Estudos e Projetos, Fundacao Carlos Chagas Filho de Amparo, Financiadora de Estudos e Projetos, Fundacao Carlos Chagas Filho de Amparo a Pesquisa do Estado do Rio de Janeiro, Conselho Nacional de Desenvolvimento Cientifico e Tecnologico and the Ministerio da Ciencia, Tecnologia e Inovacao, the Deutsche Forschungsgemeinschaft and the Collaborating Institutions in the Dark Energy Survey. The Collaborating Institutions are Argonne National Laboratory, the University of California at Santa Cruz, the University of Cambridge, Centro de Investigaciones Energeticas, Medioambientales y Tecnologicas-Madrid, the University of Chicago, University College London, the DES-Brazil Consortium, the University of Edinburgh, the Eidgenossische Technische Hochschule (ETH) Zurich, Fermi National Accelerator Laboratory, the University of Illinois at Urbana-Champaign, the Institut de Ciencies de l'Espai (IEEC/CSIC), the Institut de Fisica d'Altes Energies, Lawrence Berkeley National Laboratory, the Ludwig-Maximilians Universitat Munchen and the associated Excellence Cluster Universe, the University of Michigan, the National Optical Astronomy Observatory, the University of Nottingham, the Ohio State University, the University of Pennsylvania, the University of Portsmouth, SLAC National Accelerator Laboratory, Stanford University, the University of Sussex, and Texas A\&M University.

BASS is a key project of the Telescope Access Program (TAP), which has been funded by the National Astronomical Observatories of China, the Chinese Academy of Sciences (the Strategic Priority Research Program "The Emergence of Cosmological Structures" Grant \# XDB09000000), and the Special Fund for Astronomy from the Ministry of Finance. The BASS is also supported by the External Cooperation Program of Chinese Academy of Sciences (Grant \# 114A11KYSB20160057), and Chinese National Natural Science Foundation (Grant \# 11433005).

The Legacy Survey team makes use of data products from the Near-Earth Object Wide-field Infrared Survey Explorer (NEOWISE), which is a project of the Jet Propulsion Laboratory/California Institute of Technology. NEOWISE is funded by the National Aeronautics and Space Administration.

The Legacy Surveys imaging of the DESI footprint is supported by the Director, Office of Science, Office of High Energy Physics of the U.S. Department of Energy under Contract No. DE-AC02-05CH1123, by the National Energy Research Scientific Computing Center, a DOE Office of Science User Facility under the same contract; and by the U.S. National Science Foundation, Division of Astronomical Sciences under Contract No. AST-0950945 to NOAO.

\section{Data Availability}
The redshift catalogue data underlying this article are available in the article and in its online supplementary material. The 2D spectra data underlying this article will be made publicly available in the future with the help of the MMT data repository. Until then, the 2D spectra data will be will be shared on reasonable request to the corresponding author.

\bibliographystyle{mnras}
\bibliography{ref}
\appendix
\section{Redshift catalogue of 1054 ELGs from the MMT Binospec Pilot Study}
\label{sec:app}
\begin{table*}
    \caption{Redshift catalogue of $1054$ ELGs from this paper. The filter magnitudes are quoted from the DR8 catalogue of the Legacy Surveys. The $g$-, $r$- and $z$-band magnitudes shown here are corrected for the Milky Way extinction. The redshifts are given in barycentric frame. The confidence numbers refer to the visual confidence we attributed to the ELGs; blue and red masks refer to the fact that each of the ELGs were observed twice, once in blue and once in red gratings and as a result, each ELGs was visually inspected up to two times. Here we show the first $5$ rows. The error bars on the magnitudes can be found in the official DR8 catalogue. We do not include errorbars on the redshift measurement because our errorbars are much smaller than the precision we need. Finally, as mentioned in Section~\ref{sec:fluxing}, the variation of flux calibrations vectors are up to $20\%$ so the fluxes are considered to have an uncertainty of $20\%$. The statistical uncertainties are much smaller than this value which is why they are not included. The data underlying this article are available in the article and in its online supplementary material.}
    \label{tab:landscape}
    \begin{tabular}{ccccccccc}
    \hline
    RA & Dec & $g$ & $r$ & $z$ & Redshift & Flux & Confidence & Confidence\\
    deg & deg & mag & mag & mag & & $10^{-17}$ s$^{-1}$ cm$^{-2}$ & Blue mask & Red mask\\
    \hline
    150.09057 & 2.08496 & 23.31 & 22.92 & 22.19 & 0.904 & 9.6 & 3 & \\
    150.15421 & 2.09557 & 23.30 & 22.74 & 21.86 & 0.758 & 7.8 & 2 & \\
    150.14543 & 2.09606 & 23.09 & 23.01 & 21.89 & 0.726 & 9.0 & 3 & 3\\
    150.12804 & 2.10506 & 23.62 & 22.92 & 22.32 & 0.830 & 13.6 & 3 & 3\\
    150.10200 & 2.10547 & 19.84 & 20.05 & 19.70 & 1.141 & 15.2 & & 1\\
    \hline
    \end{tabular}
\end{table*}

\end{document}